\documentclass[useAMS,usenatbib]{mn2e}
\usepackage{graphicx}
\usepackage{booktabs}
\usepackage[usenames,dvipsnames,svgnames,table]{xcolor}
\usepackage{chngcntr}
\counterwithout{table}{section}
\usepackage{verbatim}
\usepackage{float}
\usepackage{pdflscape}
\usepackage{amsmath}
\begin{document}
\title[VLBI vs Radio-loud AGN Hosts] {Are the hosts of VLBI selected radio-AGN different to those of radio-loud AGN?}
\author[G.~Rees et al.]{G. A. Rees$^{1,2}$\thanks{E-mail: glen.rees@students.mq.edu.au}, R. P. Norris$^{2}$, L. R. Spitler$^{1,3}$, N. Herrera-Ruiz$^{4}$, E. Middelberg$^{4}$  \\
$^{1}$Macquarie University, Balaclava Road, Epping, NSW, 2109, Australia  \\
$^{2}$CSIRO Australia Telescope National Facility, PO Box 76, Epping, NSW, 1710, Australia \\
$^{3}$Australian Astronomical Observatories, PO Box 915 North Ryde NSW 1670, Australia \\
$^{4}$ Astronomisches Institut, Ruhr-Universität Bochum, Universitätstrasse 150, 44801 Bochum, Germany \\
}

\date{Accepted 2016 January 21. Received 2016 January 18; in original form 2015 December 24}
\pagerange{\pageref{firstpage}--\pageref{lastpage}} \pubyear{2014}

\maketitle
\label{firstpage}
\begin{abstract}
Recent studies have found that radio-AGN selected by radio-loudness show little difference in terms of their host galaxy properties when compared to non-AGN galaxies of similar stellar mass and redshift. Using new 1.4~GHz VLBI observations of the COSMOS field we find that approximately 49$\pm8$\% of high-mass (M $>$ 10$^{10.5}$ M$_{\odot}$), high luminosity (L$_{1.4}$ $>$ 10$^{24}$ W~Hz$^{-1}$) radio-AGN possess a VLBI detected counterpart. These objects show no discernible bias towards specific stellar masses, redshifts or host properties other than what is shown by the radio-AGN population in general. Radio-AGN that are detected in VLBI observations are not special, but form a representative sample of the radio-loud AGN population.

\end{abstract}

\section{Introduction}

Very Long Baseline Interferometry (VLBI) achieves angular resolutions at the milli-arcsecond level and as such traces only the most compact and intense sources of radio emission. On extra-galactic scales such compact emission can be attributed to either Active Galactic Nuclei (AGN), extreme clusters of Supernovae (SNe) or Supernova Remnants (SNRs). It is only in the local Universe and in the most active of star-forming galaxies however that SNe and SNRs can reach sufficient brightness to be detected by VLBI \citep{Kewley2000, Norris2007a}. Indeed, even the deepest VLBI observations (RMS $\sim$ 10~$\mu$Jy) struggle to detect SNe and SNRs at high redshifts (z $>$ 0.1). Above this, objects detected in such high sensitivity VLBI observations correspond to brightness temperatures of over 3 $\times$ 10$^{5}$K and are thus inconsistent with radio emission from star-formation \citep{Condon1991}. Above z $>$ 0.1, VLBI-detected sources also possess radio luminosities several times brighter than can be generated by reasonable SNR and SNe rates \citep{Kewley2000} or even by multiple simultaneous occurrences of the brightest supernovae currently known \citep{Parra2010}. As a result, VLBI observations offer an unambiguous way of identifying radio-AGN.

Many techniques for identifying radio-AGN focus on selecting objects with extreme radio luminosities or excesses emission in the radio-band when compared to other wavelengths. Radio-AGN identified in this way trace large scale jets and diffuse radio lobes associated with much larger temporal and physical scales than those selected by VLBI. Because of this it is possible that these two distinct methods of selecting radio-AGN may result in very different samples. Indeed VLBI observations tend to detect only 10-50\% of radio-loud AGN \citep[][Herrera-Ruiz et al., in prep]{Middelberg2011, Middelberg2013, Chi2013, Deller2014}. Is this because these are the earliest phases of the radio-AGN life-cycle or are they perhaps special in some other way? 

To test VLBI selected AGN against radio-loud AGN we can either study the properties of the AGN itself or investigate the impact it has on its host galaxy. In \citet{Rees2016} we found that the likelihood of a galaxy hosting a radio-loud selected AGN increases strongly with increasing mass. At a given mass, ellipticals and star-forming galaxies are equally likely to be radio-loud, and the optical/IR properties of radio-loud galaxies are indistinguishable from those of radio-quiet galaxies in the redshift range 0.25 $<$ z $<$ 2.25. So do VLBI-detected radio-AGN hosts follow the same evolutionary trends as their radio-loud counterparts? 

In this paper we analyse the host galaxy properties of VLBI-detected and radio-loud selected AGN in the COSMOS (Cosmological Evolution Survey) field and compare the resulting samples across a broad redshift range in terms of their host star-formation rate, color, infrared-AGN detection rate, stellar mass, dust content and radio-loudness.

\section{Data}
\label{Data}

\begin{figure}
	\centering
	\includegraphics[width=1.0\columnwidth]{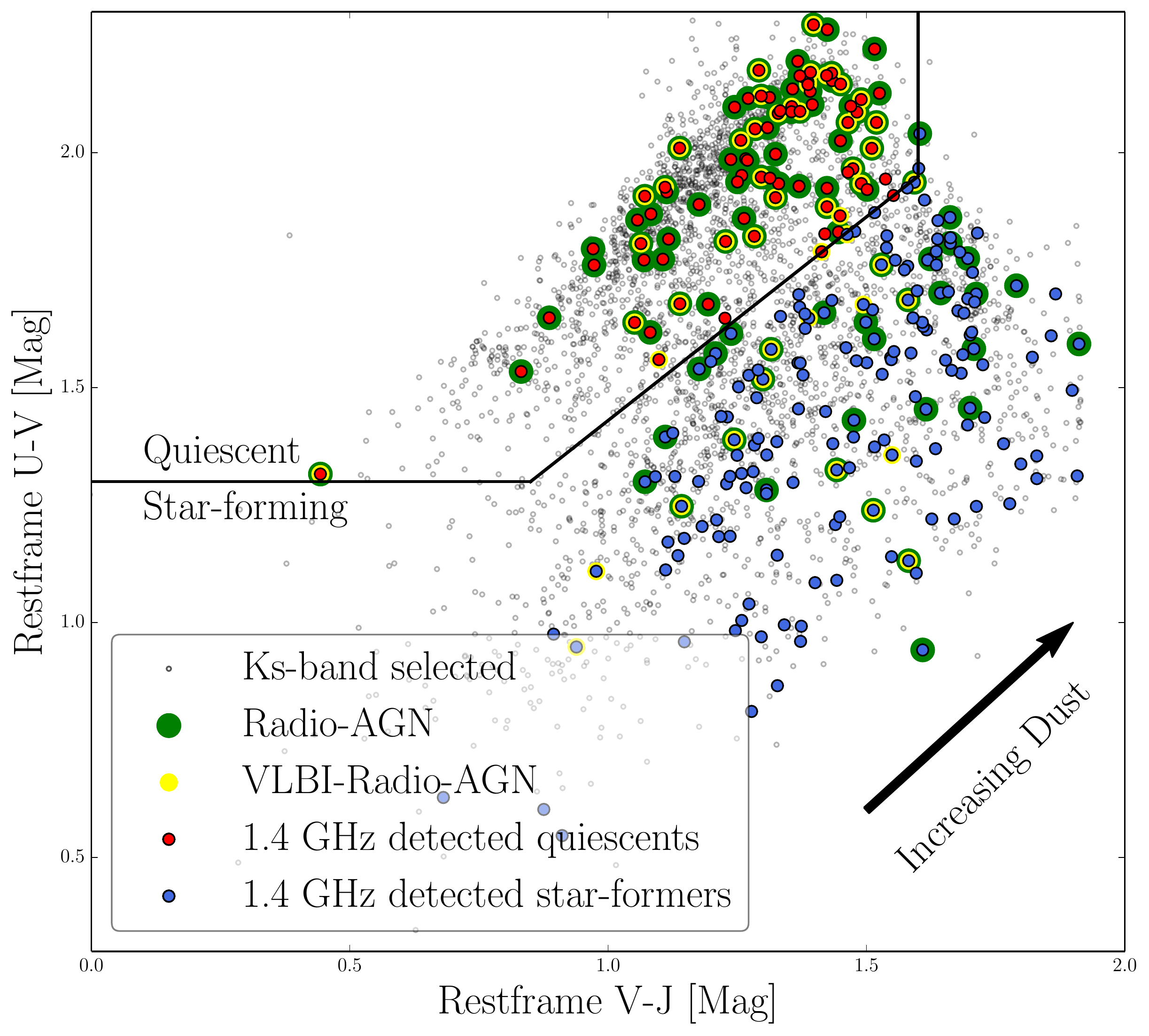}
    \caption{A rest-frame, UVJ colour-colour diagram for radio-detected star-formers (blue circles), radio-detected quiescent galaxies (red circles) and the NMBS {\it Ks}-band mass-limited sample (black circles). Radio-loud AGN are highlighted (large green circles) and those selected based on VLBI detections are also shown (large yellow circles). For clarity only 10\% of the radio non-detected sources are plotted.}
	\label{fig:UVJ-AGN}
\end{figure}

Our primary infrared data set is the Newfirm Medium Band Survey (NMBS, \citealt{Whitaker2011}) COSMOS field \citep{Giacconi2001, Schinnerer2004}, which covers an area of 27.6~$\arcmin$ $\times$ 27.6~$\arcmin$ down to a 5-sigma total magnitude of 23.5 in {\it K}--band. Using these observations along with a large amount of ancillary data, NMBS produces high quality photometric redshifts ($\Delta$z $\sim$ 1-2\%), stellar-age, stellar-mass, dust-content, star-formation rates for approximately 24,000 galaxies in the COSMOS Deep field. Using Figure \ref{fig:UVJ-AGN} we split this sample into Quiescent or Star-forming hosts using the classifications of \citet{Wuyts2007}.

Low resolution radio data for the field is taken from the 1.4~GHz Very Large Array (VLA) Cosmos Deep Project \citep{Schinnerer2010} which has a central RMS of 10~$\mu$Jy per beam. The COSMOS Deep Project covers the all of the NMBS {\it Ks/K}-band observations with an angular resolution of 2.5~$\arcsec$ $\times$ 2.5~$\arcsec$. 

Our high resolution 1.4~GHz radio data consists of deep targeted milli-arcsecond VLBI observations of each of the 2865 radio sources in the COSMOS field (Herrera-Ruiz et al., in prep). These observations were carried out on the Very Long Baseline Array and were designed to match both the maximum depth of the of the VLA Cosmos Deep Project (10~$/mu$Jy per beam) and the variation in this parameter across the field. The radio-IR identifications, in the area of overlap between the VLBI and NMBS data, are taken from \citet{Rees2016}. The resulting sample contains 385 VLA sources with NMBS counterparts and of these 64 have VLBI-detected emission. 

We further categorise all our radio sources into either radio-loud AGN or radio detected star-forming galaxies based upon their radio based star-formation rate (radio-SFR) versus a combined ultraviolet and bolometric infrared star-formation rate (UV+IR-SFR) (Figure \ref{fig:RAGN}).

\section{Analysis}
	\label{Sec:Analysis}
	The result of this process is a large sample of galaxies with information on mass, SFR, dust content, redshifts and host type which can be flagged to select radio-AGN by either radio-loudness or by VLBI detection. 

\subsection{Defining the analysis samples}
To compare the VLBI and radio-loudness selection techniques we create several different sub-samples. Our ``Sensitivity Limited sample" is limited to sources with VLA fluxes $>$ 50$\mu$Jy to achieve uniform sensitivity across the field. 

Combining the requirements of the ``Sensitivity Limited" sample with further limits on stellar mass (M $>$ 10$^{10.5}$ M$_{\odot}$), redshift (0.25 $\le$ z $\le$ 2.25) and VLA radio luminosity (L$_{1.4}$ $\ge$ 2$\times$10$^{24}$ WHz$^{-1}$) in order to have a complete sample of radio objects within our redshift range, produces our ``High-Mass/High-Luminosity" (HM/HL) sample of radio-AGN. 

We also produce a ``Mass-limited" sample of non-AGN which are simply those in NMBS above 10$^{10.5}$ M$_{\odot}$ and note that this is above the 90\% NMBS mass completeness limit out to z=2.20 \citep{Wake2011}.

Finally we build the ``mass similar" control sample by randomly sampling NMBS detected galaxies of similar redshift (same bin) and mass ($\pm$0.1 M$_{\odot}$) for each VLBI-detected radio-AGN. The median value of the control samples properties is then measured. This process is repeated 1000 times to probe the range of values seen in the control sample and the median of these runs is then plotted with the standard deviation between the 1000 runs shown as the associated error. A full description of this process can be found in \citet{Rees2016}.

\begin{figure*}
	\centering
	\includegraphics[width=1.0\columnwidth]{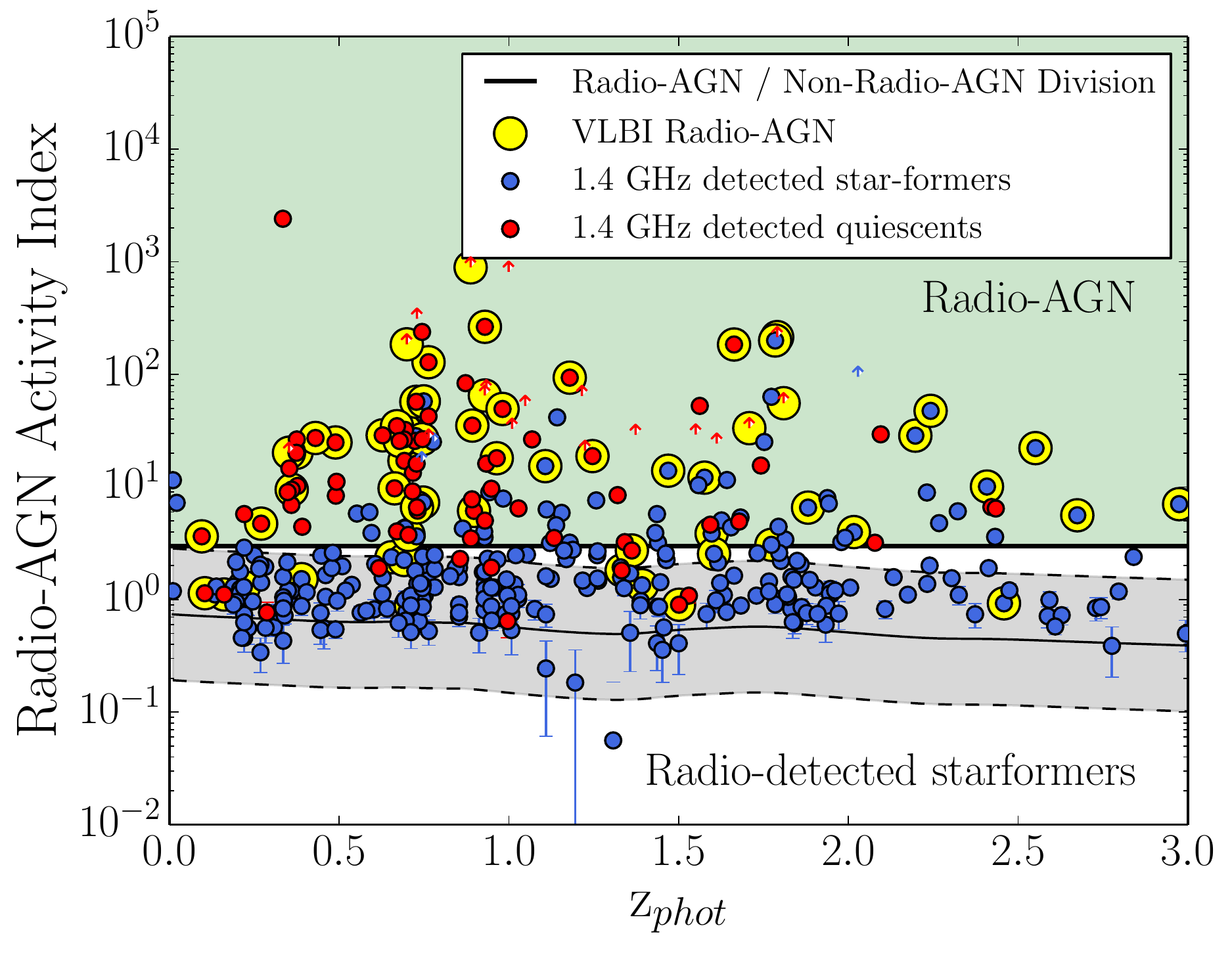}
	\includegraphics[width=1.0\columnwidth]{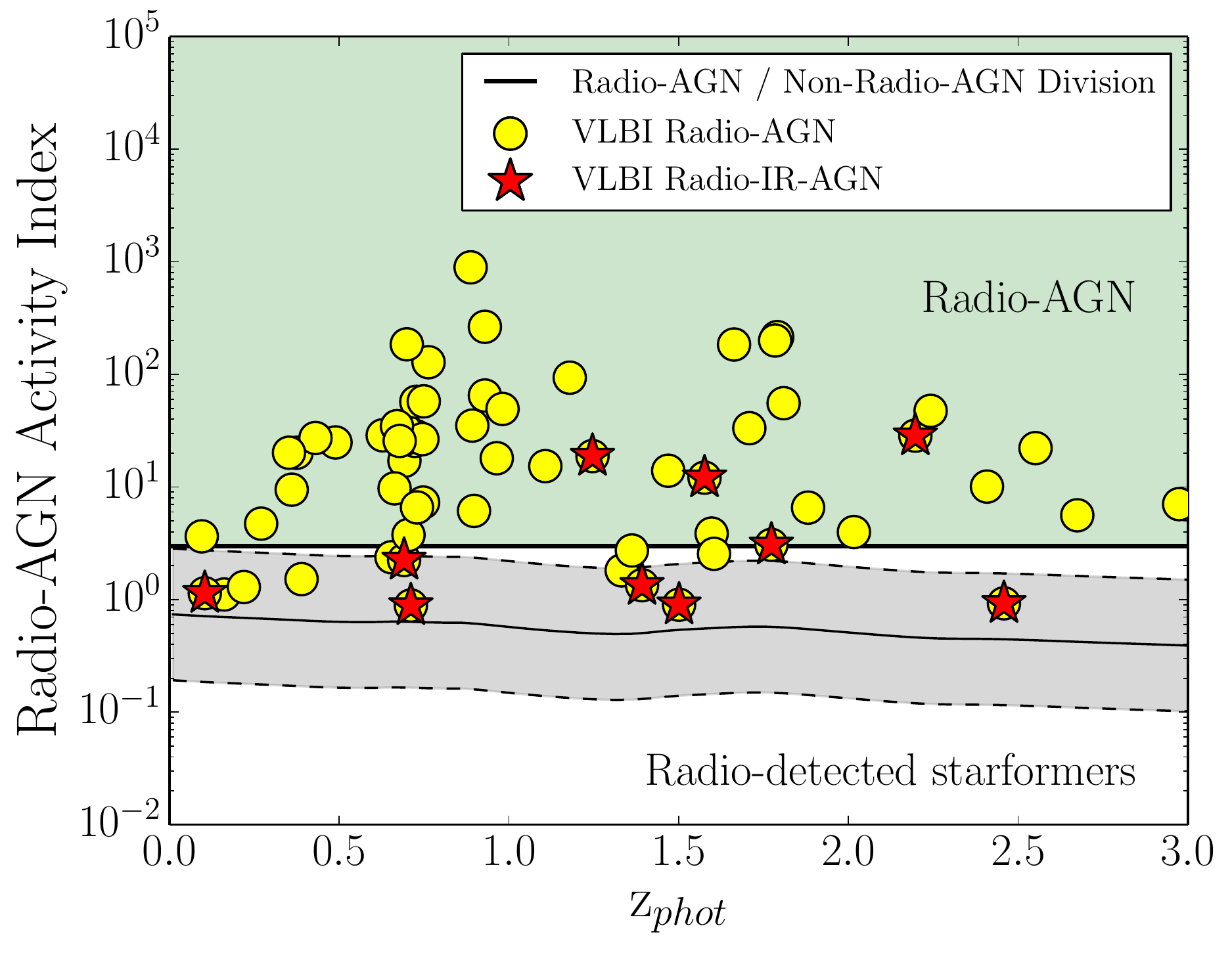}\\
	\caption{The Radio-AGN Activity index: This is simply the ratio of radio based SFR \citep{Bell2003, Karim2011} to UV+IR based SFRs \citep{Bell2005, KennicuttJr.1998}. Both of these SFRs are normalized to a Chabrier initial mass function \citep{Chabrier2003} and radio-SFR is calculated using a radio spectral index $\alpha =$ -0.3 (where S $\propto \nu^{\alpha}$) in order to match the spectral index of the radio end of the \citet{Wuyts2011} template used for calculating UV+IR SFR for star-forming objects. Objects with a radio-SFR more than 3 times higher than their UV+IR SFR (5 sigma above the Wuyts average star-former template) are classified as radio-loud AGN (green shading). Objects identified as Radio-AGN by VLBI detection are highlighted (yellow circles). We see good agreement between the techniques with the majority of VLBI sources. Right: VLBI objects below the radio-loud line are radio-quiet and of these 13 sources, 6 (46$^{+13}_{-12}$\%) posses infrared colors identifying them as infrared-AGN using the \citet{Donley2012} wedge (red stars).}
	\label{fig:RAGN}
\end{figure*}

\subsection{Are VLBI selected radio-AGN special?}
\begin{figure*}
	\centering
	\includegraphics[width=1.5\columnwidth]{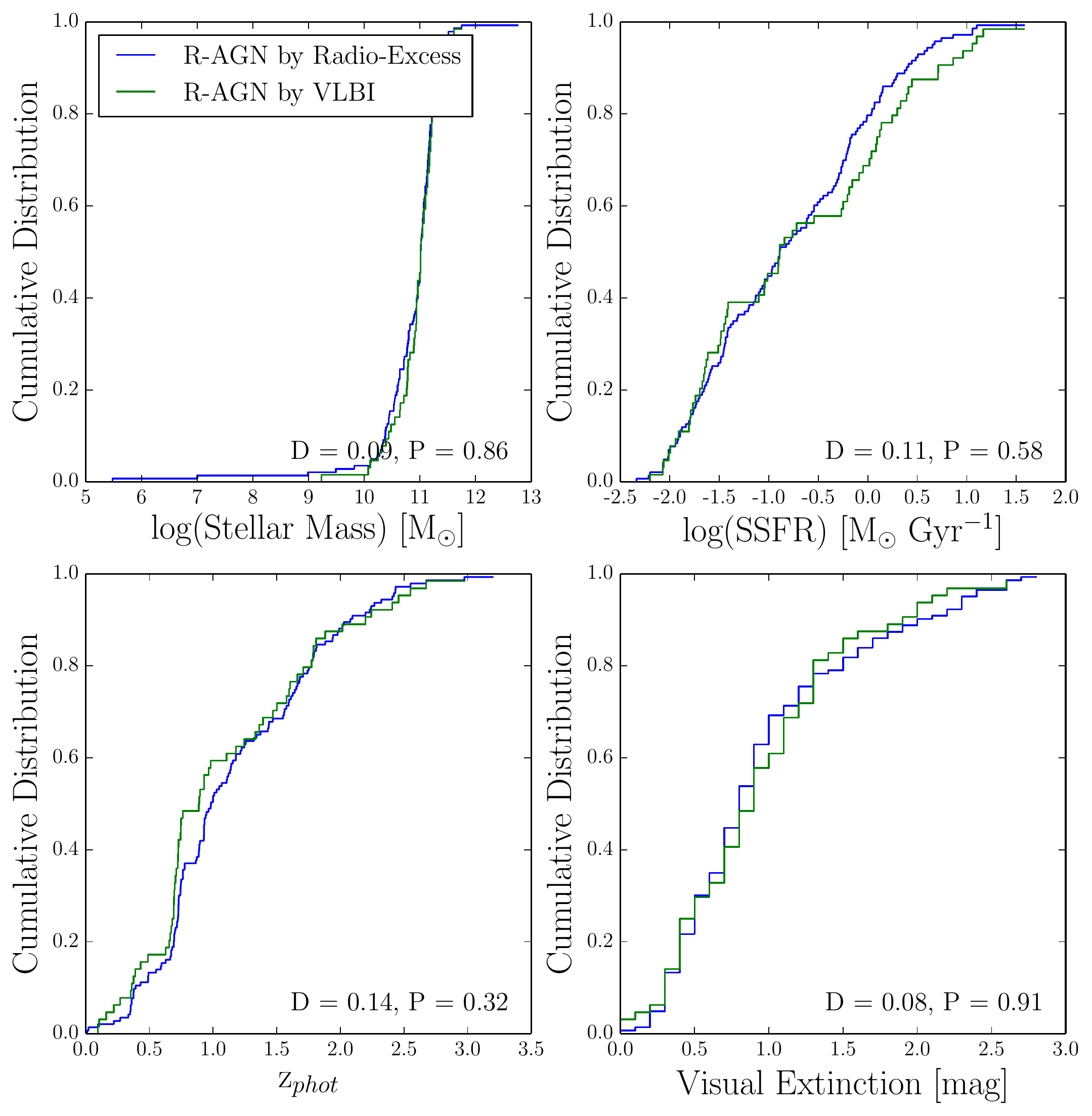}
	\caption{The Cumulative Distribution Functions for VLA detections above 50$\mu$Jy, split by VLBI and radio-loud selected AGN, against stellar mass, SSFR, redshift and visual extinction. We find no significant differences between the two populations using a Ks-test where significant is considered to be P values $<$ 0.10 and note that  this similarity continues to hold when comparing only the HM/HL radio-AGN population.}
	\label{fig:Culm}
\end{figure*}

Figure \ref{fig:Culm} shows the Cumulative Distribution Functions for our VLBI selected and radio-loud selected sensitivity limited samples in terms of stellar mass, specific star-formation rate (SSFR), redshift and dust content (as visual extinction). KS-testing these two populations shows no significant differences between them with P values ranging from P = 0.32 and up (Table 1). Table 1 also shows the KS test statistics when comparing the HM/HL VLBI sample to both HM/HL radio-loud AGN and the mass-similar sample. Even in this mass and radio luminosity complete regime we see little difference between VLBI selected radio-AGN and the radio-loud AGN or mass-similar samples.

\begin{table*}
\label{Tab:Sens}
\centering
\begin{tabular}{@{}cccc@{}}
\multicolumn{4}{c}{{\bf Table 1.}}\\
\hline
         & All: VLBI vs Radio-Loud & HM/HL: VLBI vs Radio-Loud & HM/HL VLBI vs Mass-similar \\
\hline  
P Value (Mass, SSFR, Z, Av) & 0.86, 0.58, 0.32, 0.91  & 0.99, 0.86, 1.00, 0.77     &  0.98 0.20 0.98 0.36      \\
D Value (Mass, SSFR, Z, Av) & 0.09, 0.11, 0.14, 0.08  & 0.10, 0.16, 0.10, 0.17     &  0.13 0.30 0.13 0.26      \\
Numbers (VLBI, Radio-Loud) & 64, 143                  & 21, 37                     &  21, 1217                 \\
\hline
\end{tabular}
\caption{A comparison of the stellar mass, specific star-formation rate redshift and dust-contents (through visual extinction) of our three samples using the KS-test statistics. Column 1 compares the VLBI-selected against the radio-loud selected AGN samples (both sensitivity-limited). Column 2 applies a luminosity and mass cut-off to both samples to ensure they are complete across our redshift range. Column 3 uses the same luminosity and mass cut-off for the VLBI sample and compares it against the mass and redshift-similar non-AGN population. We find that in all three cases, the samples are indistinguishable from each other.}
\end{table*}

Overall we find that VLBI detections are present for 51/143 (36$\pm$4\%) \footnote{All percentage and fraction errors in this letter are 1-sigma values calculated using the beta confidence interval as described by \citet{Cameron2011}.} of our sensitivity limited radio-loud AGN population and 18/37 (49$\pm$8\%) of the HM/HL radio-loud AGN population. Combining this with the result above suggests that VLBI selects a representative sample of radio-loud AGN and that studies using either selection method should be largely comparable. 

It is also possible to study where the radio-loud selection techniques fail. In Figure \ref{fig:RAGN} we can see that for the sensitivity limited sample, 13/64 VLBI-detected radio-AGN are missed by the radio-loudness identifier. Limiting to the high-mass, high-luminosity sample we find that this radio-quiet fraction is considerably lower with only (3/21) VLBI sources failing the radio-loudness criteria. This is in good agreement with the predicted incompleteness of the same sample in \citet{Rees2016} who estimated that 12$^{+6}_{-3}$\% of their radio-loud AGN to appear radio-quiet due to simultaneous infrared emission from the AGN which causes an artificial increase in the UV+IR SFR.              

So how many of our radio-quiet VLBI sources contain IR-AGN? Using the Donley wedge \citep{Donley2012} to identify AGN by infrared color we find that 6/13 (46$^{+13}_{-12}$\%) of the sensitivity limited radio-quiet VLBI sources are identified as IR-AGN. Correspondingly in the high-mass, high luminosity sample, this value increases to 2/3 or 66$^{+15}_{-28}$\%. For both of these samples the median ratio of VLBI to VLA flux is 54$\pm$8\%. This supports the idea that near and far infrared emission from AGN activity can result in a significant number of radio-loud AGN being mis-classified as radio-quiet or radio star-forming galaxies.

\subsection{Comparing VLBI-detected radio-AGN, radio-loud AGN and the mass similar sample}

\begin{figure*}
	\centering
	\includegraphics[width=2.0\columnwidth]{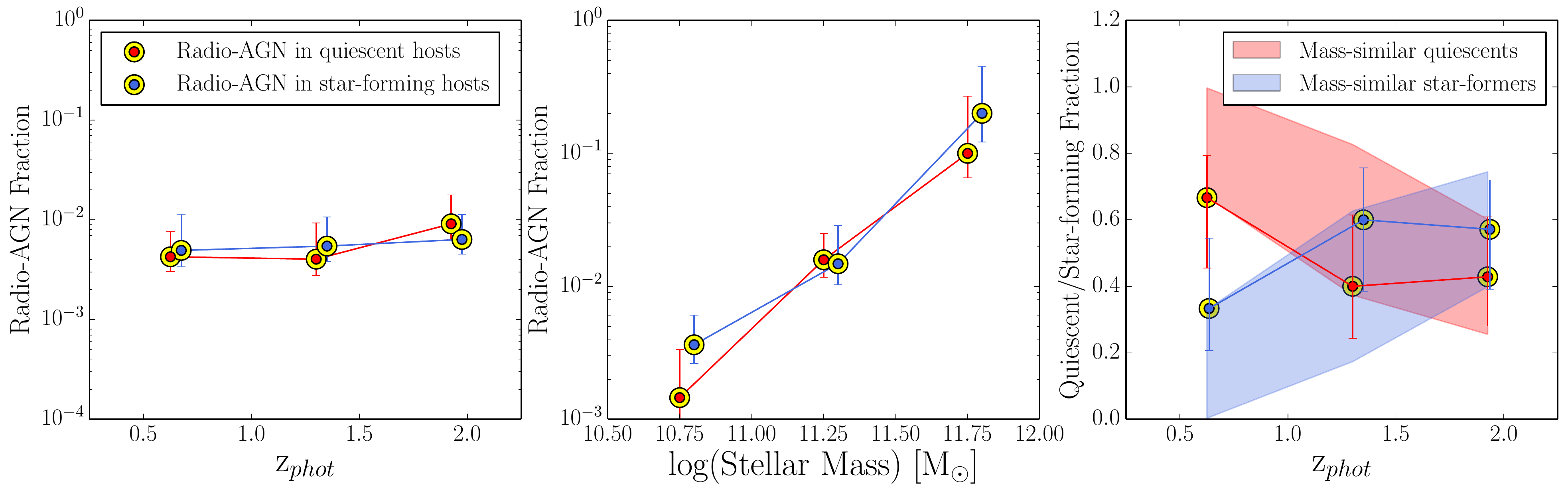}
	\caption{HM/HL VLBI detection fractions. Left: Detection fractions broken up into star-forming (blue circles) and quiescent (red circles) hosts. Essentially this is the number of HM/HL VLBI-detected sources divided by the number of NMBS sources above 10$^{10.5}$ M$_{\odot}$ as a function of redshift. Center: Same as left panel but as a function of mass. Right: Here we plot the fraction of each host type as a function of redshift for HM/HL VLBI-detected sources and the mass similar sample.}
	\label{fig:AGNFrac}
\end{figure*}
	
Our HM/HL VLBI selected AGN samples are shown in Figure \ref{fig:AGNFrac}. This is simply the number of HM/HL VLBI-detected sources divided by the number of mass-limited objects, as a function of redshift (left panel) and stellar-mass (center panel). We see no evolution in terms of redshift for the range provided and a correlation with stellar-mass, both trends are in good agreement with the those seen previous work on the radio-loud AGN population \citep{Best2005, Simpson2013, Rees2016}. Lastly, evolution of VLBI radio-AGN host types (Quiescent or Star-Forming as defined in Figure \ref{fig:UVJ-AGN}) are shown to be in reasonable agreement with the evolution of their mass and redshift similar control sample (right panel).  

\section{Discussion}
The idea that VLBI selected sources and radio-loud sources are highly similar is interesting as we might reasonably expect to identify some differences between the samples. Radio-AGN selected by radio-loudness preferentially picks out objects with relatively high radio emission for their IR emission. VLBI on the other hand selects samples with high surface brightness emission from compact sources. Despite this, we see no indication that the host galaxies of VLBI scale radio-AGN (F$_{1.4}$ $\ge$ 50$\mu$Jy, $\theta$ $\sim$ 1 mas) form a special subset of radio-loud AGN.

One obvious way to explain this result is the idea, proposed by \cite{Hickox2014}, that each active period of AGN activity is too short to directly correlate with changes in their host properties. Under this scenario, radio-AGN move periodically from active to inactive states far more rapidly than they can impact their host. As such, our sample of VLBI-detected radio-AGN are simply a random sub-sample of radio-loud AGN and hence would show little difference in their host galaxy properties. 

Finally, we find that $~$10\% of VLBI-detected sources are classified as radio-quiet despite having on average more than 50\% of their radio flux being emitted on VLBI scales. This implies that even though we might expect star-formation to provide the majority of radio emission in radio-quiet AGN below 100~$\mu$Jy \citep{Padovani2011, Bonzini2013}, a significant fraction of the radio emission from radio-quiet AGN above 50~$\mu$Jy is in fact from the AGN core. Furthermore, a high fraction of these VLBI-detected radio-quiet AGN show strong signs of AGN-based emission in the infrared, acting as a mask for their strong radio-emission.

\section{Conclusions}
\label{Sec:Conclusion}

We find that: 
\begin{enumerate}
\item VLBI selected sources account for 36$\pm$4\% of radio-loud AGN detected above 50$\mu$Jy. For our complete sample comprised of high mass, high radio luminosity sources, this fraction increases to 49$\pm$8\%. 
\item Importantly while VLBI-detected radio-loud AGN only account for a moderate proportion of the overall radio-loud AGN sample their hosts are statistically representative of the whole radio-loud population. 
\item Specifically VLBI selected radio-loud AGN hosts possess stellar-masses, SSFRs, redshifts and dust contents indistinguishable from the hosts of radio-loud AGN. 
\item This means that studies conducted using VLBI-selected samples should be in good agreement with those performed on radio-loud selected samples. 
\item Finally we find that 20$^{+6}_{-4}$\% of VLBI-detected sources are radio-quiet (14$^{+11}_{-4}$\% for the high mass, high radio luminosity VLBI-detected sample). This implies that the radio luminosity from some radio-quiet AGN can not be attributed primarily to star-formation activity.

\end{enumerate}

\label{lastpage}

This work made use of data taken on the National Radio Astronomy Observatory's Karl G. Jansky Very Large Array. The NRAO is a facility of the National Science Foundation, operated under cooperative agreement by Associated Universities, Inc

\bibliography{library}{}
\bibliographystyle{mn2e}
\appendix

\end{document}